%% file: paper.tex
\documentclass[9pt,journal,compsoc]{IEEEtran}

\usepackage[normalem]{ulem}
\usepackage{booktabs}
\usepackage{tabularx}
\usepackage{multirow}
\usepackage{caption}
\usepackage{enumitem}
\usepackage{xfrac}
\usepackage{xcolor}
\usepackage{subfig}
\usepackage{color}
\usepackage{comment}
\usepackage{psfrag}
\usepackage{graphicx,epstopdf}
\usepackage{url}
\usepackage[ruled,vlined]{algorithm2e}
\usepackage{adjustbox}
\SetAlFnt{\small}

\begin{document}
\title{Partitioning Compute Units in CNN Acceleration for Statistical Memory Traffic Shaping}

\author{
  Daejin~Jung,
  Sunjung~Lee,
  Wonjong~Rhee,\emph{~Fellow,~IEEE},
  and~Jung~Ho~Ahn,\emph{~Senior~Member,~IEEE}\\
  Department of Transdisciplinary Studies, Seoul National University
  \thanks{\copyright 2017 IEEE. This is the author supplied version of the paper which appears at IEEE Computer Architecture Letters (DOI: 10.1109/LCA.2017.2773055).}
}

\IEEEtitleabstractindextext{
\begin{abstract}

Convolutional Neural Networks (CNNs) have become the default choice 
for processing visual information, and the design complexity of CNNs 
has been steadily increasing to improve accuracy.
To cope with the massive amount of computation
needed for such complex CNNs, 
the latest solutions utilize blocking of an image over the available dimensions 
(e.g., horizontal, vertical, channel, and kernel) and batching of multiple input images
to improve data reuse in the memory hierarchy. 
While there has been a large collection of works on maximizing data reuse,
only a few studies have focused on the memory bottleneck problem caused by
limited bandwidth.
Bandwidth bottleneck can easily occur in CNN acceleration 
as CNN layers have different sizes with varying computation needs and 
as batching is typically performed over each layer of CNN
for an ideal data reuse. 
In this case, the data transfer demand for a layer can be relatively 
low or high compared to the computation requirement of the layer, and 
therefore temporal fluctuations in memory access can be induced eventually
causing bandwidth problems.  
In this paper, we first show that 
there exists a high degree of fluctuation
in memory access to computation ratio depending on CNN layers and 
functions in the layer being processed by the compute units (cores), 
where the compute units are tightly synchronized to maximize data reuse. 
Then we propose a strategy of partitioning the compute units 
where the cores within each partition process a batch of input data in a 
synchronous manner to maximize data reuse but different partitions
run asynchronously.
Because the partitions stay asynchronous and typically process different CNN 
layers at any given moment, the memory access traffic sizes of the partitions 
become statistically shuffled. 
Thus, the partitioning of compute units and asynchronous use of them 
make the total memory access traffic size be smoothened over time, and the 
degree of partitioning determines a tradeoff between data reuse efficiency
and memory bandwidth utilization efficiency.
We call this smoothing statistical memory traffic shaping, and 
we show that it can lead to 8.0\% of performance gain on a commercial
64-core processor when running ResNet-50.
\end{abstract}}

\maketitle

\IEEEdisplaynontitleabstractindextext

\input{introduction}

\input{background}

\input{contribution}

\input{experimentalsetup}

\input{evaluation}

\input{conclusion}

\section*{Acknowledgments}
This work was partially supported by the National Research
Foundation of Korea grant funded by the Korea government
(NRF-2017R1A2B2005416 and NRF-2017R1E1A1A03070560).

\bibliographystyle{IEEEtranS}
\bibliography{references}

\end{document}

%% file: introduction.tex
\section{Introduction}
\label{sec:introduction}

Emerging Convolutional Neural Network (CNN)~\cite{deep-learning-textbook}
is one of the most popular machine learning methods, especially for image
classification and object detection.
A typical CNN has a deep structure and multiple types of filters
to be able to model complicated functions, 
necessitating a high computational power.
As most of CNN's operations allow parallelism, a CNN can be fit well into
the existing data-parallel architectures, such as GPGPUs~\cite{nvidia-volta},
manycore processors~\cite{ieeemicro-2016-knl}, FPGAs~\cite{micro-2016-catapult2},
and emerging deep-learning accelerators~\cite{isca-2017-tpu}.

These architectures (called CNN accelerators hereafter) block each image over dimensions
(e.g., horizontal, vertical, channel, and kernel) and batch multiple 
images~\cite{fccm-2017-escher} to maximally exploit memory hierarchy where
components closer to compute units are smaller in capacity but offer
higher bandwidth and energy efficiency.
A large body of work has addressed maximizing data reuse 
heuristically or systematically.
For instance, Yang et al.~\cite{arxiv-2016-systematic} proposed an optimal loop blocking 
and reordering technique for convolution and fully-connected layers to
maximize data reuse in the memory hierarchy.
Their analytical model optimizes memory traffic on multi-level memory
hierarchy at a given on-chip storage budget.

Modern CNN models (architectures) such as Inception-v4~\cite{arxiv-2016-inception-v4}
and ResNet~\cite{cvpr-2016-resnet} tend to have a large number of layers.
Because the layers have different designs with a varying number of channels, number of 
kernels, and size of convolution filters, the degree of data reuse also  
varies across the layers. 
The variation can be significant, leading to a severe temporal fluctuation in bandwidth 
demands for different layers, especially for off-chip main memory.
Such a fluctuation is not a problem when the memory bandwidth is sufficiently large, 
but that is not true as we will show in Section~\ref{sec:evaluation}. 
Furthermore, memory bandwidth demand per accelerator is steadily increasing
because more transistors are becoming available due to the finer-pitch process 
technology, and because circuit- and microarchitecture-level optimizations 
(e.g., mixed-precision) to arithmetic units are allowing more arithmetic 
units to be integrated~\cite{nvidia-volta}.

Providing a sufficient main-memory bandwidth to accommodate peak demands can be
a solution. However, it is extremely inefficient as increasing main-memory
bandwidth accompanies area/energy overhead due to bulky I/Os and deteriorated
signal integrity, leading to a high-cost premium~\cite{sc-2014-microbank}.
Rather, it is desired to shrink the gap between peak and average main-memory
bandwidth utilization by regulating bandwidth demands from the numerous
compute units.
To the best of our knowledge, previous studies have not focused on this bandwidth
bottleneck problem of memory hierarchy that is caused by temporal fluctuations of resource
demands.

In this paper, we first show that the gain of data reuse
by batching diminishes on the latest CNN models, because they tend to have 
lean (smaller kernels and fewer neurons) and deep (more layers) structures. 
On the other hand, the synchronous nature of batching exacerbates 
the bandwidth fluctuation issue.
To address this problem, we propose to divide compute cores into multiple
partitions and make each partition internally operate synchronously but 
make different partitions operate asynchronously.
This solution slightly sacrifices the degree of data reuse, but its
temporal bandwidth balancing through statistical memory traffic 
shaping~\cite{computer-network-textbook} will be shown to outweigh the sacrifice.

%% file: background.tex
\begin{figure*}[!tb]
  \center
\minipage{0.61\textwidth}
  \includegraphics[width=\linewidth]{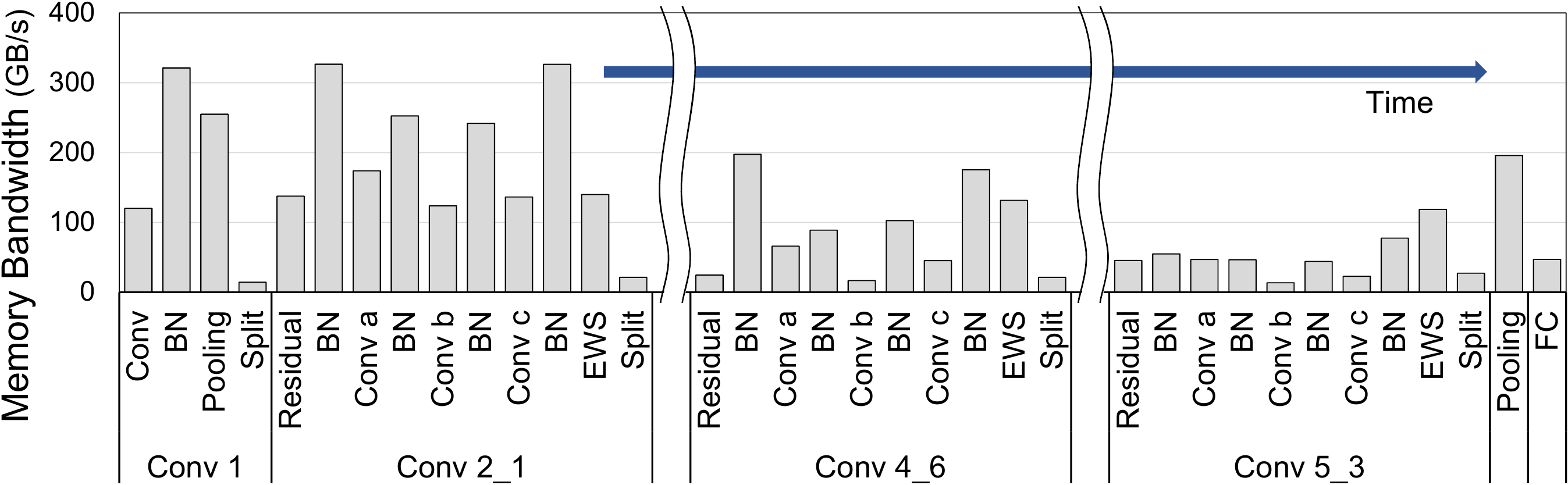}
  \caption{\small Memory bandwidth utilization on ResNet-50 CNN layers over time~\cite{cvpr-2016-resnet}.
  Convolutional layers are interleaved with other filter types (e.g., batch
  normalization (BN) and split functions), each exhibiting different bandwidth demands.}
  
  \label{fig:bw-over-time}
\endminipage
\hspace{0.19in}
\minipage{0.30\textwidth}
  \includegraphics[width=\linewidth]{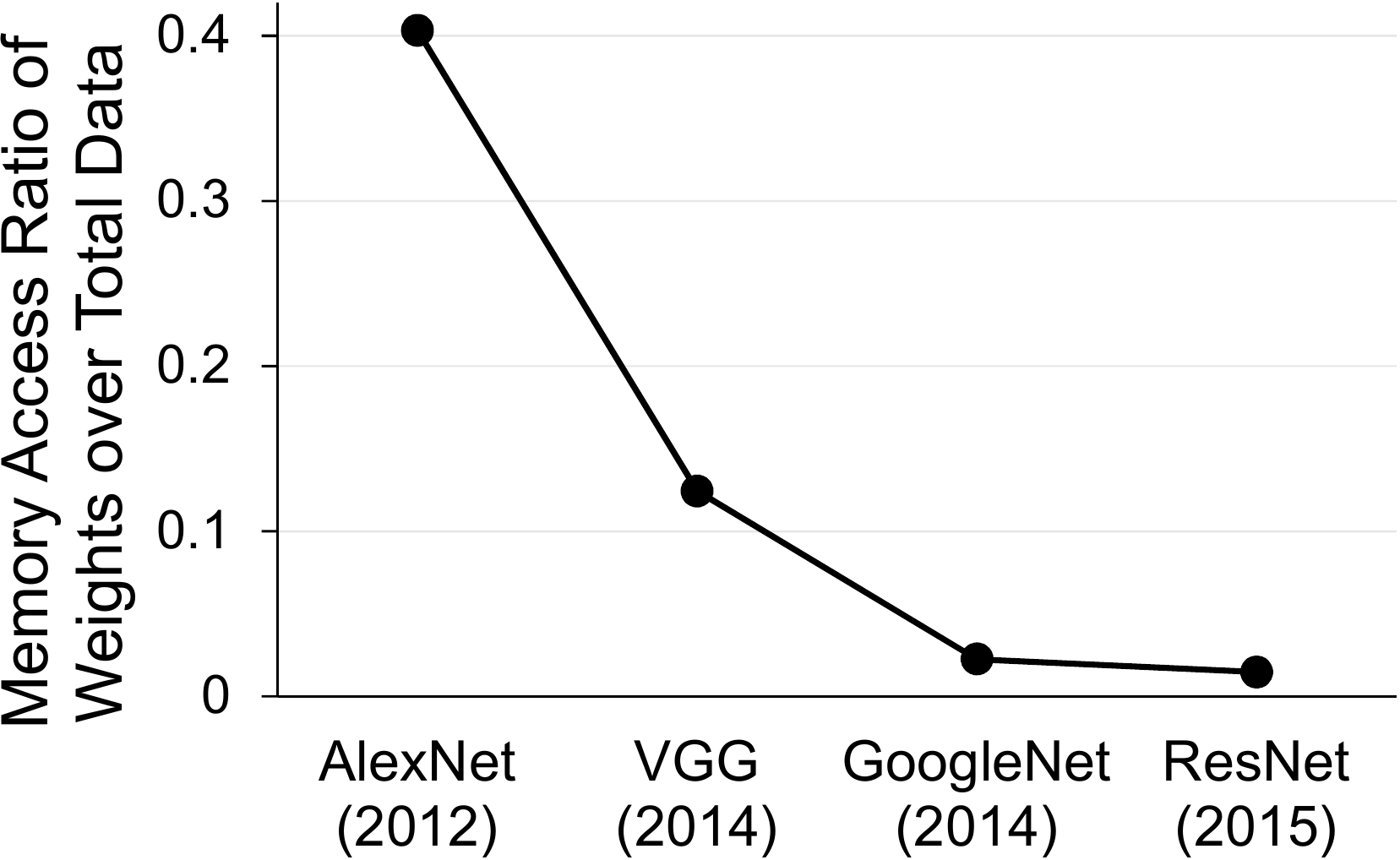}
  \caption{\small Memory access ratio of weights over total data transfers of convolutional and fully-connected layers.}
  \label{fig:weight-to-total}
\endminipage
\end{figure*}

\section{Data Reuse Characteristics of CNN}
\label{sec:datareuse}

Modern data-parallel architectures provide increasing computation performance
for CNN processing.
The latest NVIDIA GPUs based on Pascal microarchitecture reach 10 TFLOPS
for single-precision operations per chip~\cite{ieeemicro-2017-nvidia-pascal}.
Intel's Knights Landing (KNL~\cite{ieeemicro-2016-knl}) manycore CPUs
offer 6 TFLOPS per chip with 72 x86 cores featuring AVX-512 SIMD support.
As lower-precision arithmetic is viable especially for inference, CNN
accelerators including GPUs, CPUs, and FPGAs are expected to populate more
functional units tailored to 16- and 8-bit operations, further increasing
compute capability. 
For example, Google's Tensor Processing Unit~\cite{isca-2017-tpu} supports
8-bit integer performance of 92 TOPS through 64K matrix-multiply units.

Fetching data from a large memory such as an off-chip main memory per arithmetic
operation is costly in terms of energy and bandwidth efficiencies. Therefore, it is
critical to effectively utilize a hierarchy of memory components (e.g., L1,
L2 caches or scratchpad memory) by exploiting locality. 
A convolutional layer consists of \texttt{K} kernels, and each kernel receives
\texttt{C} feature maps (or input channels) passed from the previous layer, 
conducts convolution, and produces an output feature map to pass to the 
next layer.
The \texttt{K} feature maps produced in this layer are further processed by other
filters, such as pooling, rectified linear unit (ReLU), and batch normalization (BN).
The convolution filtering, which dominates CNN computation, can be made 
highly parallel and provides abundant opportunities for reusing the 
fetched weight data.
For example, the number of operations using a point in an input channel is
proportional to the number of kernels and their size
whereas a weight in a kernel is accessed for the number of times that is proportional
to the number of channels and the feature map size per channel.
Furthermore, a kernel weight can be reused further through processing input
images in batch (called batching~\cite{fccm-2017-escher} hereafter).

By making sub-blocks of feature maps and kernel weights and then by conducting computation on these
sub-blocks, we can achieve a high memory locality in reusing the weight data.
For example, the size of CNN weights often reaches tens to hundreds of 
megabytes~\cite{arxiv-2014-vggnet}, far exceeding the on-chip buffer size.
Finding optimal blocking configurations maximizes data reuse at a certain
memory capacity leading to the highest ratio of computation over accesses to
the next level in the memory hierarchy.
Yang et al.~\cite{arxiv-2016-systematic} proposed a systematic approach to find
optimal loop blocking and reordering configurations for the convolution and
fully-connected layers.
The MKLDNN library~\cite{mkldnn} we utilize in this study also exploits similar
schemes.

The latest CNN models have increasing number of layers; for example,
ResNet-50~\cite{cvpr-2016-resnet} has 50 convolutional layers, which dominate
processing time, interleaved with the other aforementioned functions to improve
a recognition rate.
These convolutional layers have a wide variation in the sizes of channels/kernels
and the number of channels/kernels.
The computation to memory access ratio heavily depends on these factors; for
example, if all the weights of the kernels at a certain convolutional layer 
fit in the last-level cache, they are loaded just once while processing a
batch of input images.
This manifests as a huge diversity in main-memory bandwidth utilization over
layers (hence over time) as depicted in Figure~\ref{fig:bw-over-time}
and Table~\ref{tab:layer-bw} 
(experimental setup is specified in Section~\ref{sec:evaluation}), but
few studies have focused on the temporal fluctuation of memory bandwidth
demands and the resulting inefficiency.

The conventional strategy of maximizing data reuse over memory hierarchy
is still effective if a CNN accelerator is equipped with main memory that
can sustain the peak bandwidth, completely absorbing the temporal fluctuation
and hence its performance being unaffected.
However, increasing peak main-memory bandwidth requires a high-cost premium.
Contemporary accelerators exploit 3D stacking of memory and better interface
material such as silicon interposer~\cite{sc-2014-microbank} to increase 
memory bandwidth, but their values are still around hundreds of GB/s, (e.g.,
732GB/s for an NVIDIA GPU with HBM2~\cite{ieeemicro-2017-nvidia-pascal}),
which is much lower than the peak bandwidth demands from compute cores
with half-precision performance (over 20 TFLOPS).
With insufficient main-memory bandwidth, compute units such as cores would
be underutilized especially during the early stages (layers) of CNN processing
(Figure~\ref{fig:bw-over-time}) whereas memory stays idle while processing the
later layers, leading to suboptimal acceleration performance if all the cores
process the same layer together.
Therefore, it is critical to devise a solution that spreads memory requests
more evenly over time to reduce the gap between peak and average demands.

%% file: contribution.tex
\begin{table}[!tb]
  \centering
  \renewcommand{\tabcolsep}{2.3pt}
  \footnotesize
    \begin{tabular}{@{}cccccccc@{}}
    \toprule
    \multicolumn{1}{l}{Layer} & \begin{tabular}[c]{@{}c@{}}Input size\\(H$\times$V)\end{tabular} & \begin{tabular}[c]{@{}c@{}}\# of input\\ channels\end{tabular} & \begin{tabular}[c]{@{}c@{}}Out size\\(H$\times$V)\end{tabular} & \multicolumn{1}{c|}{\begin{tabular}[c]{@{}c@{}}Kernel\\(H$\times$V, K)\end{tabular}} & \begin{tabular}[c]{@{}c@{}}BW\\ (GB/s)\end{tabular} & \multicolumn{1}{c}{FLOPS}   \\ 
    \midrule
    \multicolumn{1}{l}{Pooling} & \multicolumn{1}{c}{112$\times$112} & \multicolumn{1}{c}{64}   &  \multicolumn{1}{c}{56$\times$56}  &\multicolumn{1}{l|}{3$\times$3, -} & \multicolumn{1}{c}{254} & \multicolumn{1}{c}{0.6T} \\ [2pt]
    \multicolumn{1}{l}{Conv2\_1a}  & \multicolumn{1}{c}{56 $\times$56}   & \multicolumn{1}{c}{64}   &  \multicolumn{1}{c}{56$\times$56}  &\multicolumn{1}{l|}{1$\times$1, 64}  & \multicolumn{1}{c}{174} & \multicolumn{1}{c}{2.9T} \\ [2pt]
    \multicolumn{1}{l}{Conv2\_2a}  & \multicolumn{1}{c}{56$\times$56}   & \multicolumn{1}{c}{256}   &  \multicolumn{1}{c}{56$\times$56}  &\multicolumn{1}{l|}{1$\times$1, 64}  & \multicolumn{1}{c}{120} & \multicolumn{1}{c}{3.0T} \\ [2pt]
    \multicolumn{1}{l}{Conv3\_2b}  & \multicolumn{1}{c}{28$\times$28}     & \multicolumn{1}{c}{128} &  \multicolumn{1}{c}{28$\times$28}    &\multicolumn{1}{l|}{3$\times$3, 128} & \multicolumn{1}{c}{55}  & \multicolumn{1}{c}{3.7T} \\ [2pt]
    \multicolumn{1}{l}{Conv4\_3a}  & \multicolumn{1}{c}{14$\times$14}   & \multicolumn{1}{c}{1024}  &  \multicolumn{1}{c}{14$\times$14}  &\multicolumn{1}{l|}{1$\times$1, 256} & \multicolumn{1}{c}{76}  & \multicolumn{1}{c}{3.0T} \\ [2pt]
    \multicolumn{1}{l}{Conv5\_3b}  & \multicolumn{1}{c}{7$\times$7}     & \multicolumn{1}{c}{512}  &  \multicolumn{1}{c}{7$\times$7}    &\multicolumn{1}{l|}{3$\times$3, 512} & \multicolumn{1}{c}{15}  & \multicolumn{1}{c}{2.2T} \\ 
    \bottomrule
    \end{tabular}
  \caption{\small Memory bandwidth and TFLOPS of various layers on ResNet-50~\cite{cvpr-2016-resnet} (the abbreviations H, V, and K stand for horizontal, vertical, and the number of kernels respectively).}
  \vspace{-0.09in}
  \label{tab:layer-bw}
\end{table}

\begin{figure}[!tb]
  \center
  \includegraphics[width=3.1in]{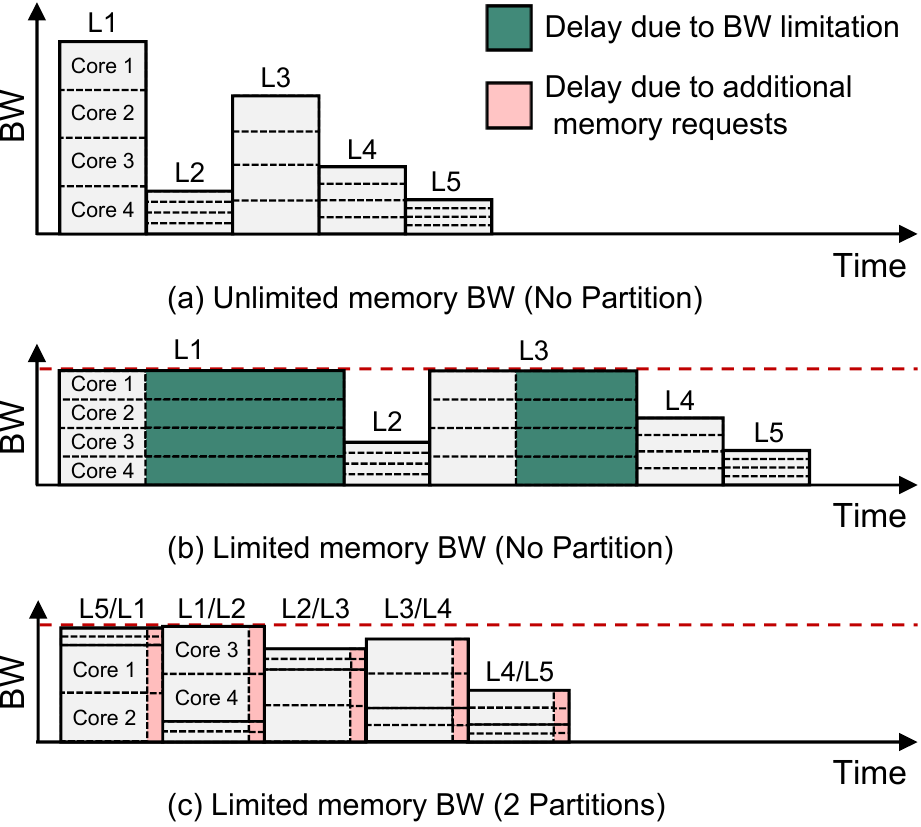}
  \caption{\small A simple illustration-purpose example of memory traffic distribution over time. When peak memory bandwidth is (a) unlimited, (b) limited, and (c) limited but using the proposed technique. 
  }
  \vspace{-0.1in}
  \label{fig:traffic-shaping}
\end{figure}

\section{Statistical Memory Traffic Shaping by Partitioning Compute Units}
\label{sec:contribution}

In this paper, we focus on alleviating the temporal fluctuation of memory
bandwidth demands on manycore-based CNN accelerators.
Similar observations and solutions can be applied to other accelerator
types supporting concurrent execution of multiple contexts (e.g., NVIDIA
Volta~\cite{nvidia-volta}).
Cores in manycore processors typically share higher levels of the memory
hierarchy (e.g., last-level caches or main memory).
This enables 1{)} certain data in a shared memory level to be used by multiple
cores (\texttt{effect$_\texttt{1}$}) and 2{)} the cores with time-varying degree of
bandwidth and capacity demands to utilize the resource more effectively
(\texttt{effect$_\texttt{2}$}).
Previous studies in CNN acceleration focused on
exploiting \texttt{effect$_\texttt{1}$} to
further enhance the degree of data reuse.
\cite{arxiv-2016-systematic} compared the options of sharing kernels or images,
and advocated sharing kernels as it can better exploit the producer-consumer
locality of images among CNN layers.
Our reference implementation~\cite{mkldnn} also shares kernels among cores;
but instead of distributing partitioned images to the cores, it allocates
different images in a batch to different cores.

This, however, tightly couples a group of cores sharing kernel weights.
In CNN acceleration, layer processing is highly sequential as a layer
takes input channels that are produced by its immediately previous
layer.
As more cores participate in a group, compute to memory access ratio increases
due to a higher degree of data (weights) reuse, but the cores operate highly
synchronously as they process the same layer together.
In an extreme case of all cores in an accelerator composing a single group, 
the configuration for data listed in Figure~\ref{fig:bw-over-time},
only a single layer is processed at any given time;
therefore \texttt{effect$_\texttt{2}$} cannot be well exploited and 
the huge variation in main-memory bandwidth demands across layers can become
a serious problem.

If accessing kernel weights takes a large portion of main-memory bandwidth
utilization, it is more beneficial to maximize kernel weight reuse by
increasing the core group size
even if it leads to more severe bandwidth fluctuation over time.
However, the impact of kernel weights on total memory bandwidth  
diminishes as CNN models advance.
Figure~\ref{fig:weight-to-total} shows
the ratio of kernel weights over total memory accesses 
for the convolutional and fully-connected
layers of the ImageNet Large Scale Visual Recognition Challenge winners.
The number of layers increases, the size of convolution filters decreases,
and a layer often receives feature maps from multiple of previously calculated layers.
These all contribute towards reducing relative portion of memory bandwidth demands
due to kernel weights.

We exploit this trend of smaller impacts of kernel weights on main-memory
accesses by separating the compute cores in a CNN accelerator into multiple
partitions.
Then we make the cores in each partition process the assigned batch
synchronously, but we allow a partition to operate asynchronously against
the other partitions.
This slightly sacrifices the degree of data reuse because kernel weights
are not shared among multiple partitions and hence should be loaded from
main memory per partition. 
However, its effect of better temporal bandwidth balancing through 
statistical memory traffic shaping~\cite{computer-network-textbook} can
outweigh the overhead.

Figure~\ref{fig:traffic-shaping} depicts the impact of this statistical
memory traffic shaping on CNN accelerator performance with an 
illustration-purpose example where memory bandwidth demands from four cores vary depending on the
layers they are processing.
With an unlimited bandwidth, the execution times of cores are not
affected by their main-memory bandwidth demands
(Figure~\ref{fig:traffic-shaping}(a)).
For a realistic system with a limited bandwidth, however, it takes much longer 
for the cores to execute layers demanding more
main-memory bandwidth (L1 and L3).
When the cores are not partitioned (Figure~\ref{fig:traffic-shaping}(b)), all four
cores should be synchronized in layer boundaries.
When the cores are divided into two partitions
(Figure~\ref{fig:traffic-shaping}(c)), the execution of core 3 and core 4
can be on different layers as the partitions operate independently.
Then, the memory bandwidth demands from the cores can be distributed such that
the aggregate bandwidth demands are always below the peak bandwidth provided
by the accelerator.
Even if there exists an additional memory traffic due to a lower degree of data
reuse in accessing kernel weights, as far as its overhead on performance 
is smaller than
the overhead due to the lack of memory traffic shaping effect,
a partitioning would be beneficial.

%% file: experimentalsetup.tex
\section{Evaluation}
\label{sec:evaluation}

\textbf{Experimental setup:}
To quantify the performance improvement, three popular CNN designs were tested: VGG-16~\cite{arxiv-2014-vggnet}, GoogleNet~\cite{cvpr-2015-googlenet}, and ResNet-50~\cite{cvpr-2016-resnet}.
The numbers of layers were chosen to be 16, 22, and 50, respectively. 
As for the processor, an Intel Knights Landing (Xeon Phi 7210) with 
64 cores was used. It has a peak single-precision arithmetic performance 
of 6 TFLOPS, and it is equipped with MCDRAM that can achieve up to 400GB/s.
We used Caffe~\cite{arxiv-2014-caffe} as the CNN framework 
and utilized Intel's math kernel library (MKL-DNN v0.1~\cite{mkldnn}) 
as it is highly optimized for Intel Knights Landing.
To measure the bandwidth utilization, hardware profiling was used.

To test the proposed strategy,
the 64 cores were divided into 2, 4, 8, and 16 partitions and 
the memory utilization and calculation speed were measured for 
each partition size. 
To keep the number of images loaded on DRAM to be constant, 
$64/n$ images were assigned to a partition as a batch where $n$ is 
the number of partitions. 
In this way, a total of 64 images are processed by the entire processor
at any time.
Because of the limitation of MCDRAM capacity (16GB), results up to 
8 partitions are provided for VGG-16, and up to 16 for GoogleNet 
and ResNet-50.
DRAM size can become the performance bottleneck for a larger number of 
partitions, but we exclude such situations because DRAM size tradeoff  
is not in the scope of this study.
Note that VGG-16's DRAM saturates faster because it needs a larger 
space for loading all of its weights.

%% file: evaluation.tex
\begin{figure}[!tb]
  \center
  \includegraphics[width=3.3in]{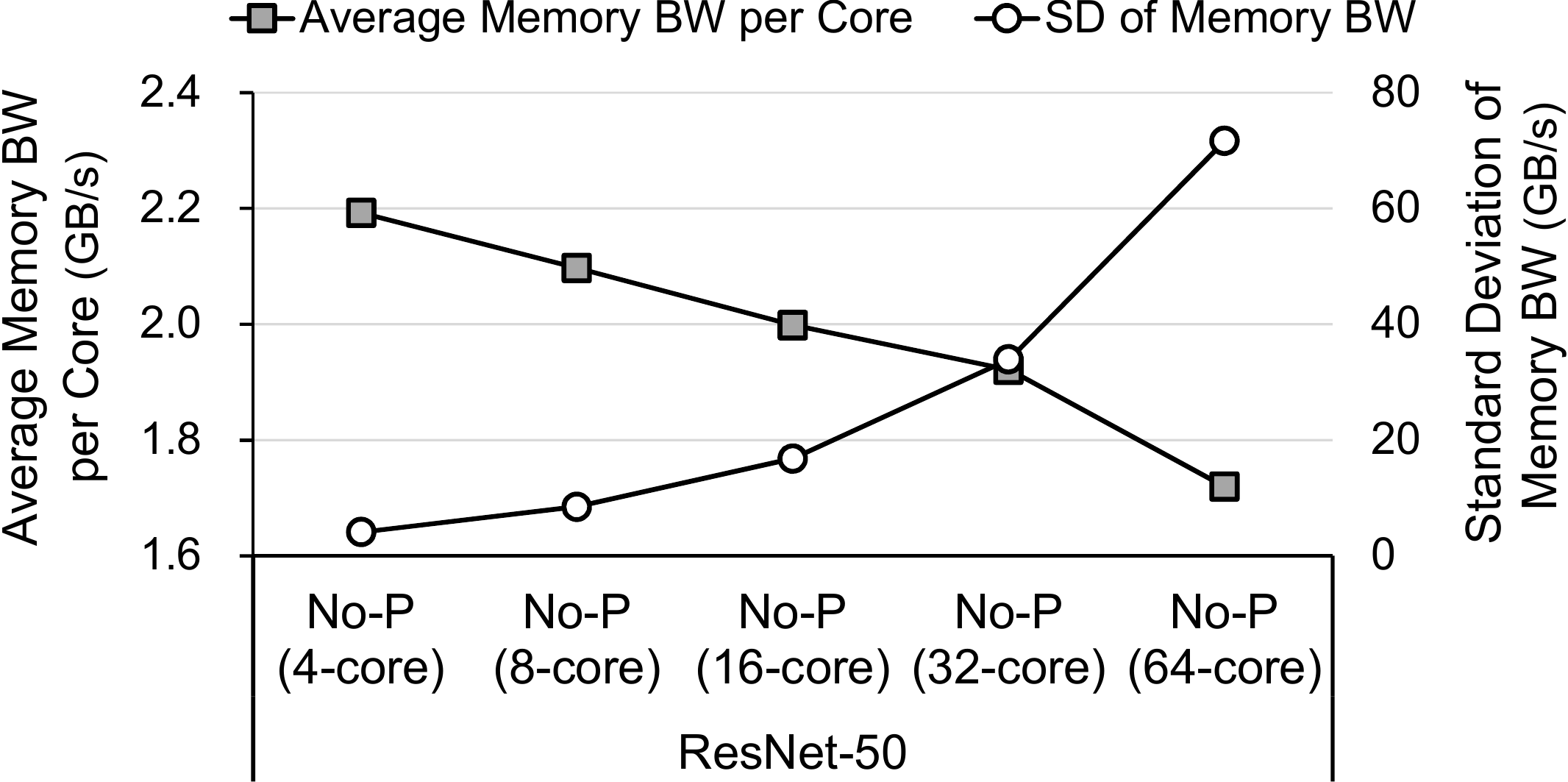}
  \caption{\small Average memory bandwidth per core and standard deviation of memory bandwidth for increasing number of cores. Plotted for ResNet-50.}
  \vspace{-0.08in}
  \label{fig:in-com-power}
\end{figure}

\begin{figure}[!tb]
  \center
  \includegraphics[width=3.3in]{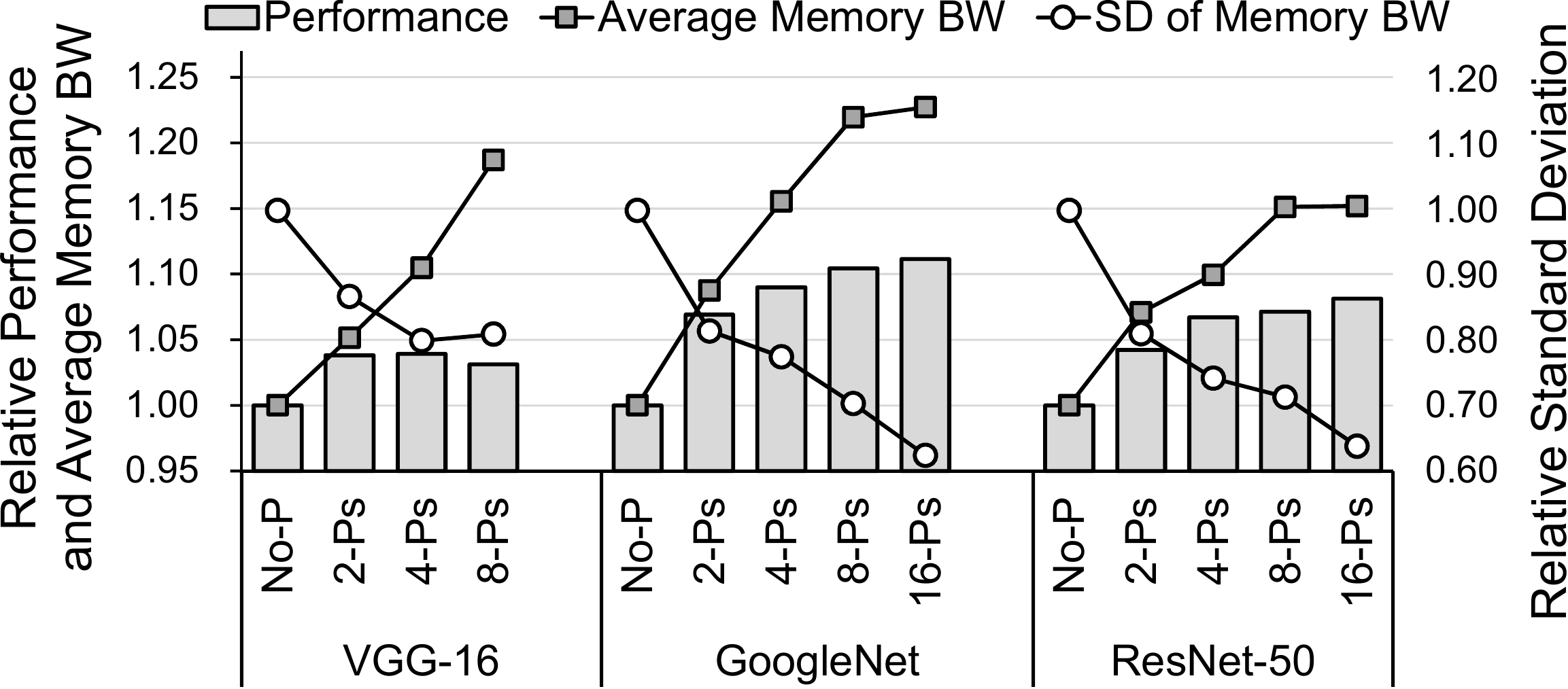}
  \caption{\small Relative performance, standard deviation of memory bandwidth, and average of memory bandwidth 
  for increased partition sizes. Shown for VGG-16, GoogleNet, and ResNet-50.}
  \vspace{-0.08in}
  \label{fig:performance}
\end{figure}

\begin{figure}[!tb]
  \center
  \includegraphics[width=3.4in]{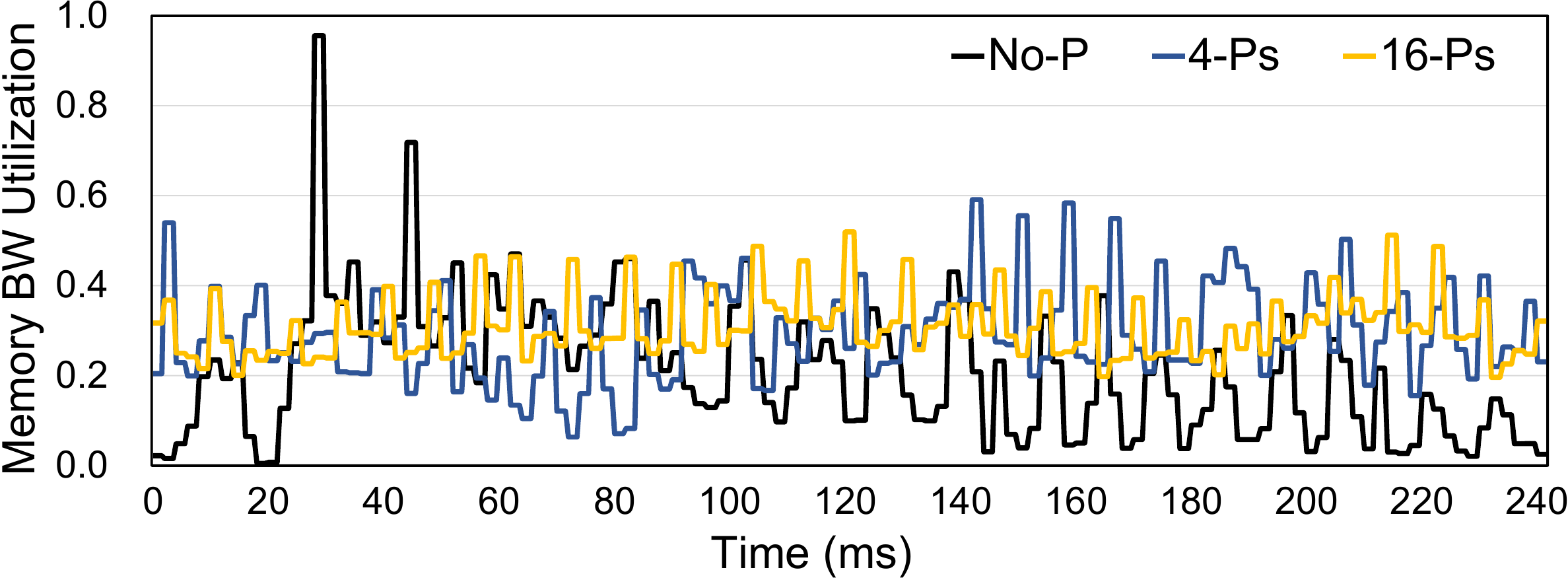}
  \caption{\small Memory bandwidth utilization over time for no-P, 4-Ps, and 16-Ps. Plotted for ResNet-50.}
  \vspace{-0.08in}
  \label{fig:mem-util-time}
\end{figure}

\textbf{Results:}
The baseline performance of synchronous data reuse is investigated first. 
In Figure~\ref{fig:in-com-power}, the average and standard deviation of memory bandwidth usage are shown for ResNet-50 with no partition. 
The image batch size for data reuse is the same as the number of cores such that each core processes a single image per weight loading from DRAM.
It can be observed that the standard deviation increases as the number of cores increases. 
This is expected because more cores are equivalent to more concurrently processed images, and therefore 
a larger fluctuation in the absolute size of total bandwidth usage (in GB/s). 
As the standard deviation becomes larger, there is a higher chance of the memory bandwidth becoming the performance bottleneck. 
This leads to a decrease in the average memory bandwidth usage per core as shown in Figure~\ref{fig:in-com-power}, because more time needs to be spent for waiting in the queue.
Note that this memory bandwidth bottleneck problem is expected to become even more crucial 
when compute capability is further improved with 16- and 8-bit operations.  

To address the memory bottleneck problem, 
we applied the proposed partitioning strategy to the three CNN models, 
and present the relative performance results in Figure~\ref{fig:performance}.  
For VGG-16, GoogleNet, and ResNet-50, standard deviation is reduced by up to 20.0\%, 37.6\%, and 36.2\%, respectively.
This confirms that the fluctuation is reduced by increasing the partition size. 
It is a direct result of statistical traffic shaping over asynchronous partitions, and thus the average bandwidth usage, i.e., memory bandwidth utilization efficiency, is also improved by 18.7\%, 22.7\%, and 15.2\%, respectively.
Eventually, the overall performance is improved by 3.9\%, 11.1\%, and 8.0\%.

The partitioning improves performance of all three CNN models. 
The performance improvement comes despite of less weight data reuse, because the bandwidth issue is more critical for the tested cases. 
For the set of chosen test scenarios, the performance is steadily improved as the number of partitions is increased except for VGG-16's 8 partitions. 
In general, we expect the performance to deteriorate as the number of partitions becomes too large, but the limitation on DRAM size prevented us from testing such scenarios.
The performance improvement is most significant when partition size is increased from 1 (no partition) to 2. 
This is because the reduction of fluctuation by traffic shaping is most significant for the case. 
Figure~\ref{fig:mem-util-time} shows the memory bandwidth utilization of no partition, 4 partitions, and 16 partitions for ResNet-50. 
Without partitioning, memory bandwidth utilization severely fluctuates.
For 16 partitions, however, the bandwidth utilization becomes relatively steady.

%% file: conclusion.tex
\section{Conclusion}

For CNN acceleration, a synchronous use of cores has been considered 
as a desirable solution because of its data reuse efficiency. 
In this work, we have shown that such a synchronous use can have a downside of 
a memory bandwidth bottleneck problem, especially for the latest 
CNN algorithms whose weight data reuse is less critical.
To provide a mechanism for trading data reuse efficiency with 
memory bandwidth utilization efficiency, we proposed a partitioning
strategy where compute units are divided into multiple partitions
and different partitions run asynchronously. 
The strategy was tested over VGG-16, GoogleNet, and ResNet-50
using Intel Knights Landing processor
with 64 cores. 
The evaluation results show that the standard deviation of memory bandwidth 
usage is reduced by 20.0-37.6\% and the average is increased by 15.2-22.7\%. 
This indicates that a statistical traffic shaping is achieved and the 
memory bandwidth is better utilized on the average. 
Overall, CNN acceleration performance is improved by 3.9-11.1\%.